\documentclass[11pt,a4paper]{amsart}

\usepackage[latin1]{inputenc}
\usepackage[english]{babel}
\usepackage{amsmath}

\usepackage{amssymb, mathabx}
\usepackage[numbers]{natbib}
\usepackage{graphicx}
\usepackage{braket}
\usepackage[pdftex,plainpages=false,colorlinks,hyperindex,bookmarksopen,linkcolor=red,citecolor=blue,urlcolor=blue]{hyperref}

\usepackage{epstopdf}

\usepackage{braket}

\bibpunct{[}{]}{;}{n}{,}{,}
\usepackage[hmargin=3cm,vmargin={3.5cm,4cm}]{geometry}

\theoremstyle{definition}                                 

\theoremstyle{definition}                           

\theoremstyle{remark}                             

\usepackage{color}

\newcommand{\be}{\begin{eqnarray}}
\newcommand{\ee}{\end{eqnarray}}

\def\eg{{\it e.g. }} 
\def\ie{{\it i.e. }}

\def\eg{{\it e.g.}\ }
\def\ie{{\it i.e.}\ }

\numberwithin{equation}{section}

\allowdisplaybreaks

\begin{document}
\title{A one parameter class of\\
	 Fractional Maxwell-like models}

    \author{Ivano Colombaro$^1$}
		\address{${}^1$ Department of Physics $\&$ Astronomy, University of 	
    	    Bologna and INFN. Via Irnerio 46, Bologna, ITALY.}
		\email{ivano.colombaro@bo.infn.it}
	
	    \author{Andrea Giusti$^2$}
		\address{${}^2$ Department of Physics $\&$ Astronomy, University of 	
    	    Bologna and INFN. Via Irnerio 46, Bologna, ITALY.}
		\email{andrea.giusti@bo.infn.it}
	
    \author{Francesco Mainardi$^3$}
    	    \address{${}^3$ Department of Physics $\&$ Astronomy, University of 	
    	    Bologna and INFN. Via Irnerio 46, Bologna, ITALY.}
			\email{francesco.mainardi@bo.infn.it}



    \date  {\today}

\begin{abstract}
In this paper we discuss a one parameter modification of the well known fractional Maxwell model of viscoelasticity. Such models appear to be particularly interesting because they describe the short time asymptotic limit of a more general class of viscoelastic models known in the literature as Bessel models.
\end{abstract}

    \maketitle

\noindent \textit{Paper presented on the occasion of the\\
2017 International Conference on Applied Mathematics and Computer Science.}

\section{Introduction}
		Linear viscoelasticity has revealed to have an insightful relevance in various branches of science, from condensed matter physics to medical physics. Its relevance has grown even further thanks to the massive development of viscoelastic models based on Fractional Calculus, mostly within the last three decades. For some historical and quantitative references, we invite the interested reader to refer to some world literature masterpieces such as \cite{Mainardi-1997, Mainardi-Spada 2011, Mainardi_BOOK10}.
	
	As it is well known in the literature (see \eg \cite{Meral-2010}), many materials appear to exhibit both elastic and viscous behavior, such as biological tissues and blood vessels (see \eg \cite{Meral-2010, AG-FM_ICMMB14}). The main idea behind the mathematical structure of viscoelastic models is that  the order of the time derivatives in the constitutive (stress-strain) equation characterizes the material's viscoelastic properties. The power of fractional calculus lies in allowing the formulation of intermediate models based on time derivatives of fractional order. These models appear to be way more appropriate if we seek a simpler and accurate description of the dynamics of certain complex systems in biophysics.

	The aim of this paper is to discuss some implications of a specific class of viscoelastic models, known as Bessel Models (see \cite{IC-AG-FM-2016}) that share the feature of showing a continuous transition from a fractional Maxwell-like model of order $1/2$, for short time, to an ordinary Maxwell model for long time.
	
	In particular, in the first Section we present a brief review of the main formalism of linear viscoelasticity. Moreover, we also briefly discuss the main mathematical properties of the so called Bessel Models.
	
	In the second Section, we discuss the underlying asymptotic fractional models corresponding to the short time asymptotic limit of the Bessel Models and how they relate to the renown fractional Maxwell model of order $1/2$.

	Finally, in the last Section we complete the paper with concluding remarks and hints for future
research.

	
\section{Linear Viscoelasticity $\&$ Bessel Models} \label{sec-2}
	As widely discussed in \cite{Mainardi_BOOK10}, a viscoelastic body
can be considered as a linear system in which the stress plays the role of excitation
function for a certain material while the strain act as the response function, or vice versa.
	Given this picture, it might seem clear how the Heaviside  $\Theta-$function $\Theta(t)$ plays a fundamental role as an input function for these systems. In the following, we denote by ${J}(t)$ the strain response to the unit step of stress, according to the {\it creep test}. Whereas, according to the {\it relaxation test}, we denote by ${G}(t)$ the stress response to a unit step of strain.

    These functions are usually referred to as the
{\it creep compliance} and {\it relaxation modulus}
respectively, or, simply, as {\it material functions}
of the viscoelastic system. Moreover, because of the the causality
condition, both functions have to vanish for $t<0$.

It is worth remarking that the limiting values of the material functions
for $t \to 0^+$ and $t \to +\infty$  are, respectively, related to the
(glass) instantaneous and equilibrium behaviors of the viscoelastic
body. As a consequence, it is common to denote
   $J_g := J(0^+) \ge 0$  the {\it glass compliance},
   $ J_e := J(+\infty)\ge 0 $ the {\it equilibrium compliance},
and
  $ G_g := G(0^+)\ge 0$  the {\it glass modulus}
  $ G_e := G(+\infty)\ge 0$   the {\it equilibrium modulus}.
Indeed, both the material functions appear to be 
non--negative with 
 $ J(t)$   {\it non decreasing} and
 $G(t)$  {\it non increasing}.

Denoting by $\sigma(t)$ and $\epsilon(t)$ the uniaxial stress and strain, respectively, and 
under the hypotheses of sufficiently well behaved causal histories,
 in most cases the  constitutive equations can be written in the following forms
\begin{equation} \label{stress-strain}
{\epsilon (t)} = J_g\, \sigma(t) +  ( {\dot J} \star \sigma) (t) \,,
\qquad
{\sigma (t)} = G_g\, \epsilon(t) +  ( {\dot G} \star \epsilon) (t) \,,
\end{equation}
where the $\star$ represents the convolution product.

	Given that the equations in Eq.~(\ref{stress-strain}) are of convolution type, they can be conveniently studied by means of the technique of Laplace transforms. Then, in the Laplace domain, they read
\begin{equation}
\widetilde \epsilon (s) = s\, \widetilde{J}(s) \, \widetilde \sigma(s)\,,
\qquad
\widetilde \sigma  (s) = s\, \widetilde{G}(s) \, \widetilde \epsilon (s)\,,
\end{equation}
from which one can infer the so called {\it reciprocity relation}
\begin{equation} \label{reciprocity}
  s\, \widetilde{J}(s)  = \frac{1}{s\,\widetilde{G}(s)}
 \,. 
 \end{equation}
Now, recalling the limiting theorems for the Laplace transform, one can easily prove that
\begin{equation} \label{limits}
J_g = {1}/ {G_g}, \quad  J_e = {1}/{G_e}\,, 
\end{equation}
with the convention that $0$ and $+\infty$ are reciprocal to each other.

Moreover, the causal functions $\dot J(t)$ and $\dot G(t)$ are referred as
the {\it rate of creep (compliance)} and
 the {\it rate of relaxation (modulus)}, respectively.
These functions play the role of {\it memory functions} in the constitutive
equations (\ref{stress-strain}).
  
Assuming  $J_g >0$  and  $G_g>0$, which means that we are focusing our attention 
to viscoelastic models exhibiting instantaneous elasticity,  it is convenient
to consider the memory functions
scaled with their corresponding initial values:
\begin{equation} \label{Psi-Phi} 
\Psi(t) =  \frac{1}{J(0^+)}\,  \frac{dJ}{dt}\,, \quad
\Phi(t) = -\frac{1}{G(0^+)}\, \frac{dG}{dt}\,.
\end{equation}
These functions are here required to be  completely monotonic (CM) functions, \ie they have to be non--negative, non-increasing functions for $t>0$ with infinitely many derivatives alternating in sign.
As outlined by several authors, see e.g. \cite{Hanyga STAMM04},
\cite{Mainardi_BOOK10}, 
the assumption of completely monotonicity is sufficient for the physical significance of the models. Indeed, such a requirement ensures, for instance, that in isolated systems the energy decays monotonically as expected from physical arguments.

	It is now worth reviewing a generalization of the fluid-like model proposed by Giusti and Mainardi in various papers, see \cite{AG-FM_MECC16, AG-FM_WSEAS14, AG-FM_ICMMB14}, based on modified Bessel functions of order $0,1,2$ in the Laplace domain. This new class of models based modified Bessel functions of contiguous order, known in the literature as {\it Bessel models}, was first introduced by Colombaro, Giusti and Mainardi in \cite{IC-AG-FM-2016}.
	
	The Bessel models are featured by the following material functions, in the Laplace domain,
	\be
	 && s \, \widetilde{J} (s; \nu) = 1 + \widetilde{\Psi} _{\nu} (s) = 
	1 + \frac{2 (\nu + 1)}{\sqrt{s}} \frac{I_{\nu + 1} (\sqrt{s})}{I_{\nu + 2} (\sqrt{s})} \, , \\
 	&& s \, \widetilde{G} (s; \nu) = 1 - \widetilde{\Phi} _{\nu} (s) =
	1 - \frac{2 (\nu + 1)}{\sqrt{s}} \frac{I_{\nu + 1} (\sqrt{s})}{I_{\nu} (\sqrt{s})} \, ,
	\ee
	for $\nu > -1$.
	
	 Now, inverting back to the time domain and taking advantage of the results in \cite{AG-FM-EPJP}, gives
\begin{equation} \label{eq_J2}
J (t; \nu) = 2 \left( \frac{\nu + 2}{\nu + 3} \right) + 4 (\nu + 1) (\nu + 2) t - 4 (\nu + 1) \sum _{n=1} ^{\infty} \frac{1}{j_{\nu + 2 , \, n} ^{2}} \exp \left( - j_{\nu + 2 , \, n} ^{2} \, t \right) \, , 
\end{equation}
\begin{equation} \label{eq_G2}
G (t; \nu) = 4 (\nu + 1) \sum _{n=1} ^{\infty} \frac{1}{j_{\nu , \, n} ^{2}} \exp \left( - j_{\nu , \, n} ^{2} \, t \right) \, .
\end{equation}
This clearly shows that the time representation of the material functions is given in terms of absolutely convergent generalized Dirichlet series, for $t>0$ and $\nu > -1$. 

	For sake of clarity and completeness, in the following we show some plots of the material functions $J (t; \nu)$ and $G (t; \nu)$ in terms of a non-dimensional time coordinate $t$.

\begin{center}
\includegraphics[width=10cm]{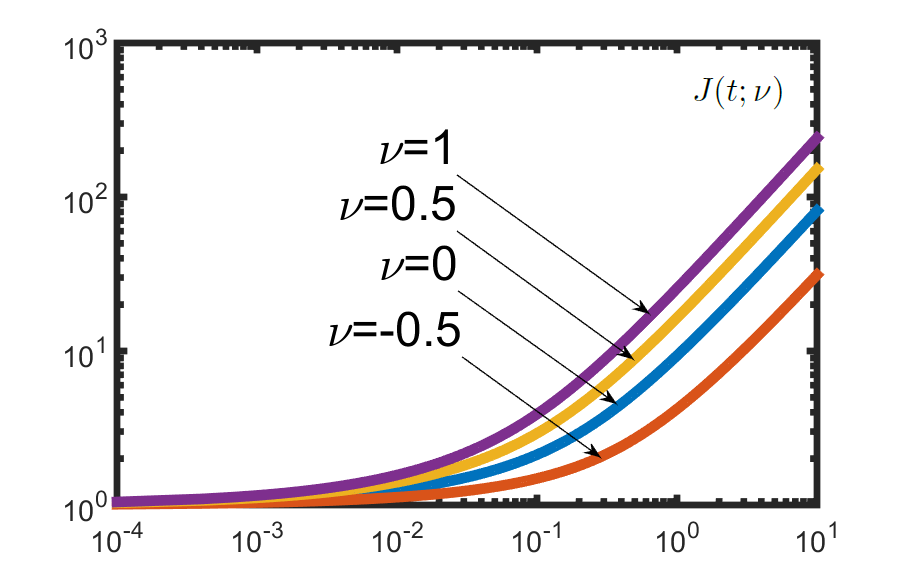}
\end{center}
Fig.1: The creep compliance  $J (t; \nu)$  for $\nu=-0.5, 0, 0.5, 1.$

\begin{center}
\includegraphics[width=10cm]{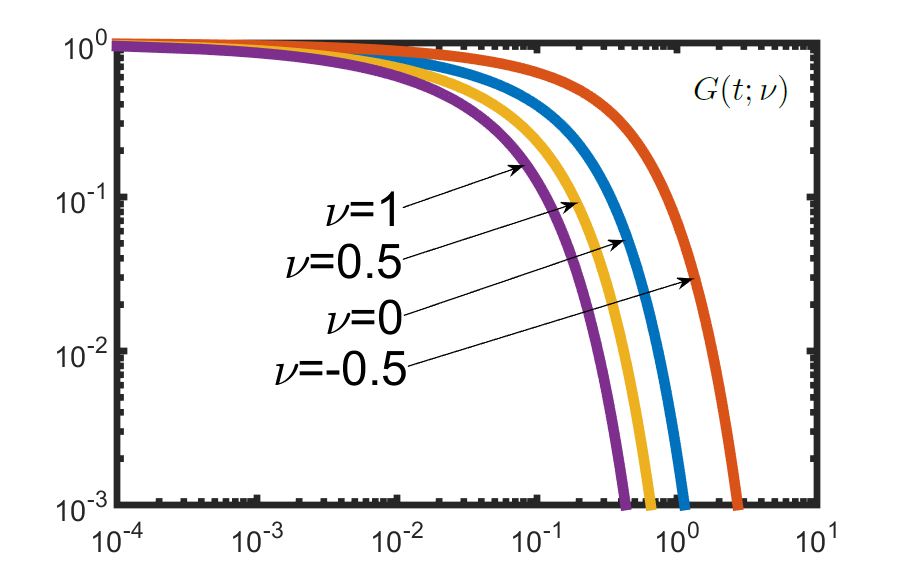}
\end{center}
Fig.2: The relaxation modulus $G (t; \nu)$  for $\nu=-0.5, 0,0.5, 1.$
\vskip 0.5truecm

For further details on the Bessel Models, we invite the interested reader to refer to \cite{IC-AG-FM-2016}.

\section{Bessel Models $\&$ Modifications of Fractional Maxwell} \label{sec-3}
	One of the most relevant fractional viscoelastic models is the renowned Fractional Maxwell model of viscoelasticity (see \eg \cite{Mainardi-Spada 2011, Mainardi_BOOK10}) of order $\alpha = 1/2$, \ie
	\begin{equation}\label{eq-max-1/2}
	\sigma (t) + a_{1} \, _{C} D _{t} ^{1/2} \sigma (t) = b_{1} \, _{C} D _{t} ^{1/2} \varepsilon (t)
	\end{equation}
	where $_{C} D _{t} ^{1/2}$ is the Caputo's fractional derivative of order $1/2$ and $a_{1}, b_{1} > 0$ are some physical coefficients connected with the relaxation time $\tau$ and to the shear parameter $\mu$, specifically $\tau ^{1/2} = b_{1} /2 \mu$ and $\mu = b_{1}/2a_{1}$.
 	
	Taking the Laplace transform of both sides of Eq.~(\ref{eq-max-1/2}), one gets
	\begin{equation}
	\left( 1 + a_{1} \, s^{1/2} \right) \ \widetilde{\sigma} (s) 
	= 
	b_{1} \, s^{1/2} \widetilde{\varepsilon} (s) \, .
	\end{equation}
	Notice that the initial conditions at $t = 0^{+}$ do not explicitly appear in the constitutive equation (\ref{eq-max-1/2}). The reason why we did not specify the initial conditions comes from the fact that, in order to guarantee the equivalence between the integral representation of stress-strain relations and  constitutive equations, the contributions from these conditions must not appear, namely they have to be vanishing or cancel in pair-balance. For further details, we suggest to refer to \cite{Mainardi_BOOK10}.
	
	Now, recalling that
	$$ \widetilde{\varepsilon} (s) = s \widetilde J (s) \ \widetilde{\sigma} (s) \, , $$
one can easily show that
\begin{equation} \label{eq-j-max}
s \widetilde J _{M} (s) = \frac{1 + a_{1} \, s^{1/2}}{b_{1} \, s^{1/2}} \, ,
\end{equation}
and, following the result in Eq.~(\ref{reciprocity}) one can immediately infer that
	\begin{equation} \label{eq-g-max}
	s \widetilde G _{M} (s) = \frac{b_{1} \, s^{1/2}}{1 + a_{1} \, s^{1/2}} \, .
	\end{equation}
If we then set $\tau = 1$ (which implies $a_{1} = 1$ and $b_{1} = 2 \mu$), for sake of simplicity, the previous expressions reduce to
\be
&& s \widetilde J _{M} (s) = \frac{1}{b_{1}} \left( 1 +  s^{-1/2} \right) \, , \\
&& s \widetilde G _{M} (s) = b_{1} \, \frac{s^{1/2}}{1 + s^{1/2}} \, .
\ee
	If we now invert the expressions (\ref{eq-j-max}) and (\ref{eq-g-max}) back to the time domain, we recover the very well known material functions of the Farctional Maxwell model of order $1/2$, \ie
	\be
	&& J _{M} (t) = \frac{a_{1}}{b_{1}} \left( 1 + \frac{2}{a_{1} \sqrt{\pi}} \, t^{1/2} \right) \, ,\\
	&& G _{M} (t) = \frac{b_{1}}{a_{1}} \, E_{1/2} \left( - \frac{t^{1/2}}{a_{1}} \right) \, ,
	\ee
where the function $E_{1/2} (z)$ is the one parameter Mittag-Leffler function of order $\alpha = 1/2$. The definition in the complex plane of this function is provided by its Taylor powers series around $z = 0$, that is
	$$ E_{\alpha} (z) = \sum _{n=0} ^{\infty} \frac{z^{n}}{\Gamma (\alpha n +1)} \, , 
	\qquad 
	\alpha > 0 \, . $$ 	
	
	For the Bessel models \cite{IC-AG-FM-2016} we have that (assuming $\tau  =1$)
\begin{equation}\label{eq-j-bessel}
s \widetilde J (s; \nu) \sim 1 +  2 (\nu + 1) \ s^{-1/2}  \, , \qquad s \to \infty \, , \quad \nu > -1 \, .
\end{equation}
	The latter shows some strong similarities with the Creep Modulus for the Fractional Maxwell model of order $1/2$, at least in the short time limit (indeed, $t \to 0^{+} \,\Leftrightarrow \, s \to \infty$ due to the Tauberian theorem for Laplace transforms). Then, it seems quite interesting to investigate the form of the constitutive equation for a viscoelastic model inspired by the material function in Eq.~(\ref{eq-j-bessel}). Let us consider, therefore, a set of viscoelastic models featuring the following one parameter creep material function in the Laplace domain,
	\begin{equation}
	s \, \widetilde J_{as} (s; \nu) = 1 +  2 (\nu + 1) \ s^{-1/2}  \, , \quad \nu > -1 \, ,	
	\end{equation}
	from which we can easily get
	\begin{equation} \label{eq-new-max}
	s \, \widetilde J_{as} (s; \nu) = \frac{2 (\nu + 1) + s^{1/2}}{s^{1/2}} \, .
	\end{equation}
Then, the constitutive equation, in the Laplace domain, takes the following form, 
\begin{equation}
\left[ 1 + \frac{s^{1/2}}{2 (\nu + 1)} \right] \ \widetilde{\sigma} (s) = \frac{s^{1/2}}{2 (\nu + 1)} \ \widetilde{\varepsilon} (s) \, .
\end{equation}	
Hence, inverting back to the time domain, one gets
\begin{equation} \label{eq-mod-model}
\left[ 1 + \frac{1}{2 (\nu + 1)} \ _{C} D _{t} ^{1/2} \right] \ \sigma (t) 
= 
\frac{1}{2 (\nu + 1)} \ _{C} D _{t} ^{1/2} \varepsilon (t) \, .
\end{equation}
	It is now important to remark that the condition $\nu > -1$ is not just a mathematical requirement rising from \cite{IC-AG-FM-2016}. Indeed, it can rather be traced back to the physical meaning (see \cite{Mainardi-Spada 2011, Mainardi_BOOK10, Hanyga STAMM04}) of the coefficients that appears in the constitutive equation of the underlying asymptotic model described in Eq.~(\ref{eq-mod-model}).
	
	Finally, starting from the Creep material function in Eq.~(\ref{eq-new-max}), one can easily recover the time representation of the material functions for this ``new'' class of (asymptotic) models, \ie
	\be 
	&& \label{eq-fin-1} J_{as} (t; \nu) = 1 + \frac{4 (\nu + 1)}{\sqrt{\pi}} \, t^{1/2} \, , \\
	&& \label{eq-fin-2} G_{as} (t; \nu) = E_{1/2} \left( - 2 (\nu + 1) \, t^{1/2} \right) \, .
	\ee
	
	\begin{center}

\includegraphics[width=10cm]{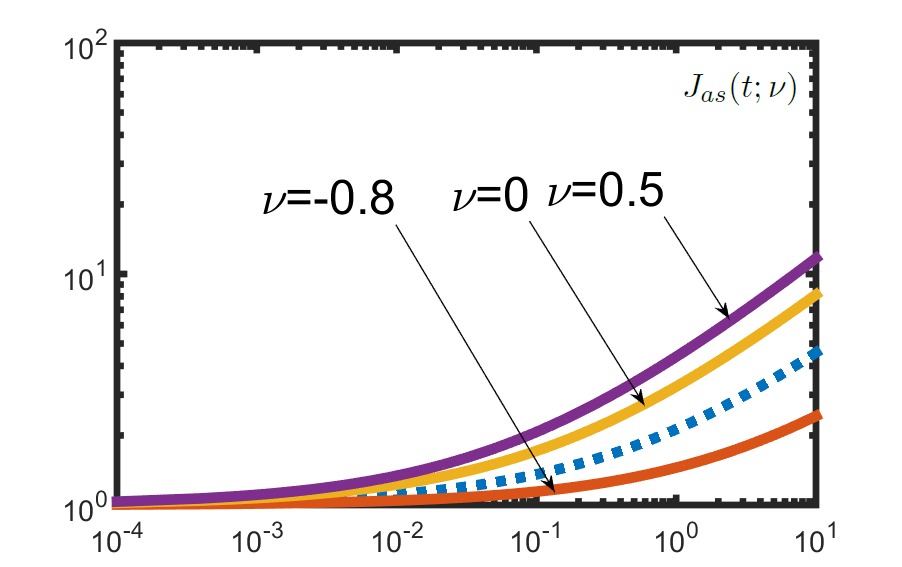}
\end{center}
\vskip 0.1truecm
Fig.3: Comparison between the creep compliance $J_{M} (t)$ of the fractional Maxwell model of order $1/2$ (dashed line) and the creep compliance $J_{as} (t; \nu)$ of the Maxwell-like class of models for $\nu=-0.8, 0, 0.5$ (continuous lines).

\vskip 1 truecm

\begin{center}
\includegraphics[width=10cm]{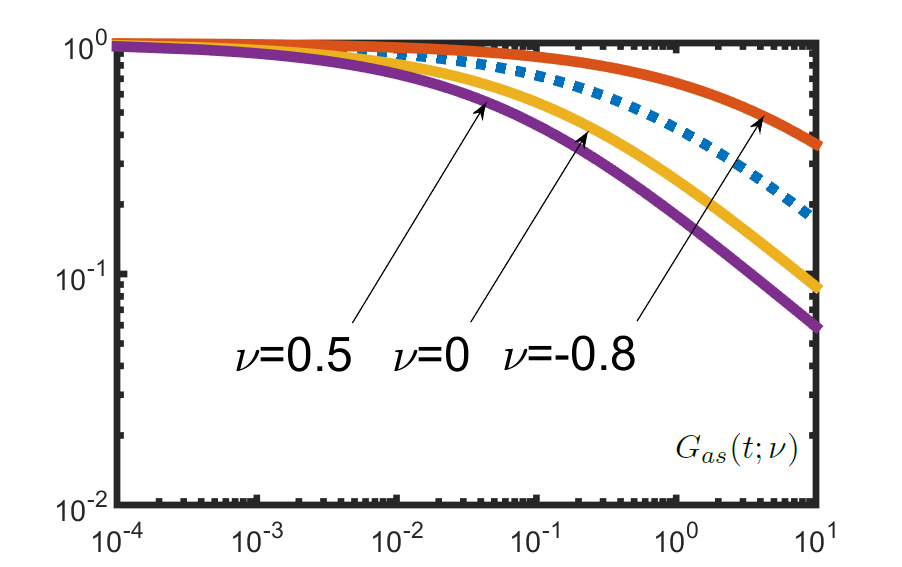}
\end{center}
\vskip 0.1truecm
Fig.4: Comparison between the relaxation modulus $G_{M} (t)$ of the fractional Maxwell model of order $1/2$ (dashed line) and the relaxation modulus $G_{as} (t; \nu)$ of the Maxwell-like class of models for $\nu=-0.8, 0, 0.5$ (continuous lines).
\vskip 0.5truecm

\newpage
	
	\section{Concluding Remarks} \label{sec-4}
	After a brief review of the basic formalism of fractional viscoelasticity, in the second Section we have provided a complete discussion of a class of viscoelastic models, parametrized by $\nu > -1$, corresponding to time asymptotic limit of a more general class of viscoelastic models known as Bessel models. In particular, we have provided the asymptotic constitutive equations as well as the corresponding material functions, both depending on $\nu > -1$. This allowed us to provide a physical justification for the condition $\nu >-1$ identifying the strict connection between this parameters the coefficients that appears in the constitutive equations of linear viscoelasticity.
	
	Moreover, just looking at the expressions in (\ref{eq-fin-1}) and (\ref{eq-fin-2}), together with Fig. 3 and Fig. 4, we can fairly state that, although these asymptotic models resemble the Fractional Maxwell model of order $1/2$, they still present a peculiar dependence on $\nu$ that is worth of further investigations. Specifically, it would be interesting to perform the asymptotic expansions for transient viscoelastic waves, discussed in \cite{Buchen-Mainardi 1975}, within the framework of Bessel models and Fractional Maxwell-like models, discussed in this paper.

\section*{Acknowledgements}
	The work of A. G. and F. M. has been carried out in the framework of the activities of the National Group of Mathematical Physics (GNFM, INdAM).
	


 \end{document}